\documentclass[prb,twocolumn,amsmath,amssymb]{revtex4}

\usepackage{color}
\usepackage{bm}
\usepackage{dcolumn}% Align table columns on decimal point
\usepackage[dvips]{graphicx}

\usepackage{ulem}

\begin{document}

\preprint{preprint}

\title{
Quantum Hall effect of Dirac fermions on the surface of a topological insulator 
}

\author{Takeshi Furusawa and Takahiro Fukui}
%\author{Takanori Fujiwara}
\affiliation{Department of Physics, Ibaraki University, Mito 310-8512, Japan}

\date{\today}

%%% abstract %%%
\begin{abstract}
We study the quantum Hall effect of Dirac fermions on the surface of a Wilson-Dirac type 
topological insulator thin film in the strong topological insulating phase.
Although a magnetic field breaks time reversal symmetry of the bulk, the surface states 
can survive even in a strong field regime.
We examine how the Landau levels of the surface states are affected by symmetry breaking 
perturbations. 
\end{abstract}

\pacs{
}

\maketitle

Graphene has led our interest to Dirac fermions in crystals.\cite{Novoselov:2005aa,Zhang:2005aa}
On the honeycomb lattice, the valence band and the conduction band  linearly touch
in the Brillouin zone, which yields massless Dirac fermions.\cite{Zheng:2002aa}
This feature is quite manifest under a strong magnetic field, in which 
unusual quantum Hall effect (QHE) for relativistic particles \cite{Ishikawa:1984aa,Ishikawa:1985uq,Semenoff:1984aa}
has been observed.\cite{Novoselov:2005aa,Zhang:2005aa}
Various topological aspects of graphene QHE such as disorder effects \cite{Sheng:2006aa}
and the bulk-edge correspondence \cite{Hatsugai:2006aa}
have been investigated.
In graphene, there appear two Dirac fermions in the Brillouin zone because of the doubling mechanism 
on lattice systems.\cite{Nielsen:1981aa,Nielsen:1981ab}
Therefore, the Hall conductivity as the result of degenerate Dirac fermions is always observed. 

Two-dimensional Dirac fermions can also be observed on the surface of 
three dimensional (3D) topological insulators.\cite{Fu:2007aa,Moore:2007ab,Roy:2009dq,Hasan:2010fk,Qi:2011kx}
Although they are doubled as well, it may be easier to control them, because 
they appear on the opposite surfaces which are spatially separated.
Recently, several experimental and theoretical studies on the QHE of the surface states 
of topological insulators have been 
reported. \cite{Yoshimi:2015aa,Yoshimi:2015ab,Zhang2017,Morimoto:2015aa}
In particular, in a magnetic topological insulator with broken inversion symmetry
nondegenerate surface states have been realized and
the QHE for a single Dirac fermion has been observed.\cite{Yoshimi:2015ab}

The surface states are ensured by topologically nontrivial phase of the bulk known as the bulk-edge
correspondence. \cite{Hatsugai:1993fk}
Therefore, it is not obvious whether
the QHE of massless Dirac fermions are indeed observed in a strong magnetic field regime,
since broken time reversal symmetry makes the bulk topological insulating phase unstable.
Motivated by this question, we investigate in this paper the stability of the surface Dirac fermions 
of a topological insulator under a strong magnetic field.

\begin{figure}[htb]
\begin{center}
\begin{tabular}{cc}
\includegraphics[scale=0.45]{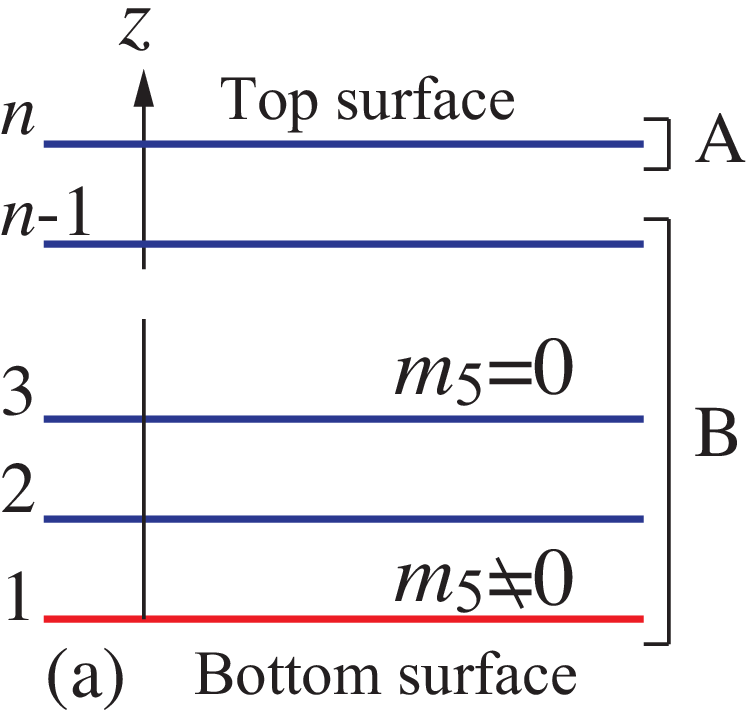}&
\includegraphics[scale=0.55]{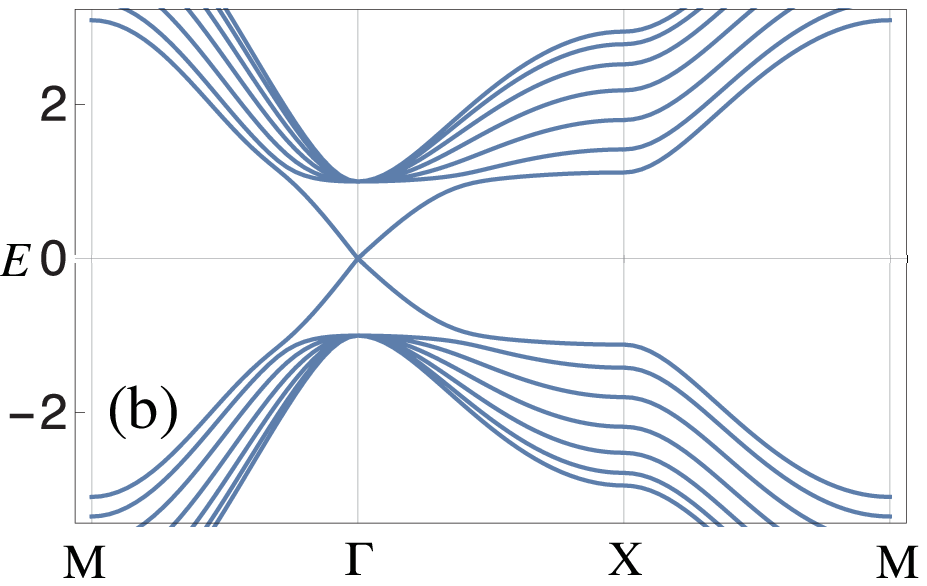}
\end{tabular}
\begin{tabular}{ccc}
\includegraphics[scale=0.295]{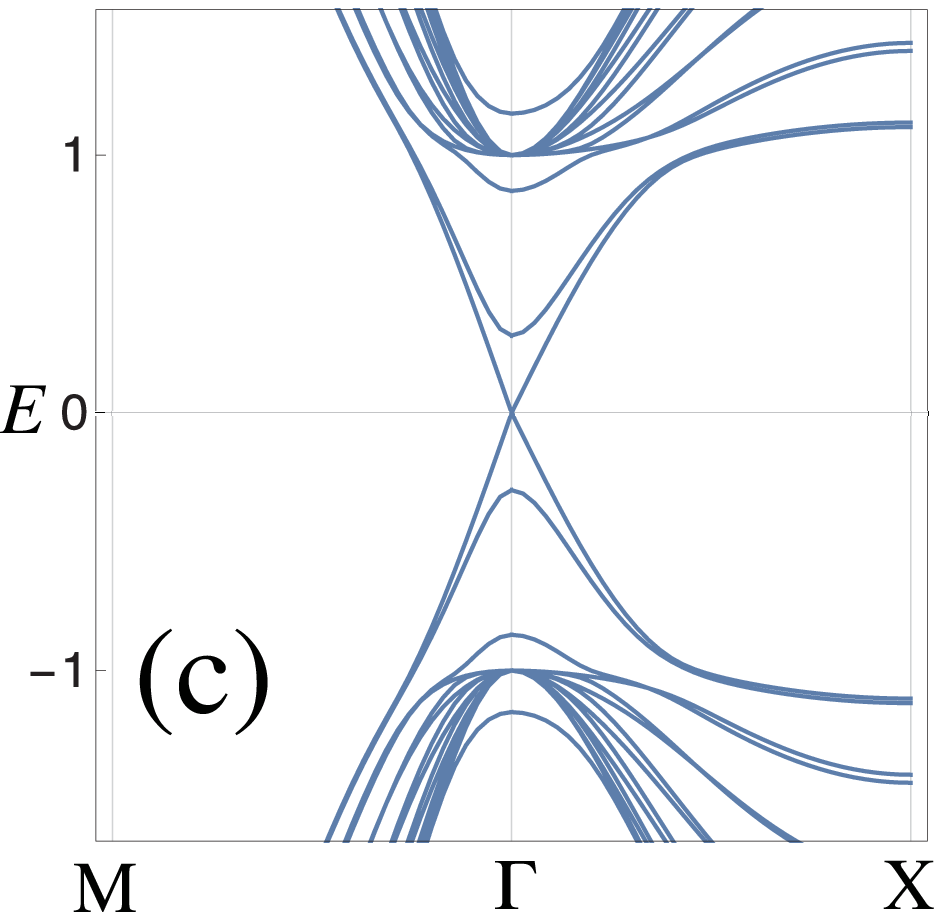}
&\includegraphics[scale=0.295]{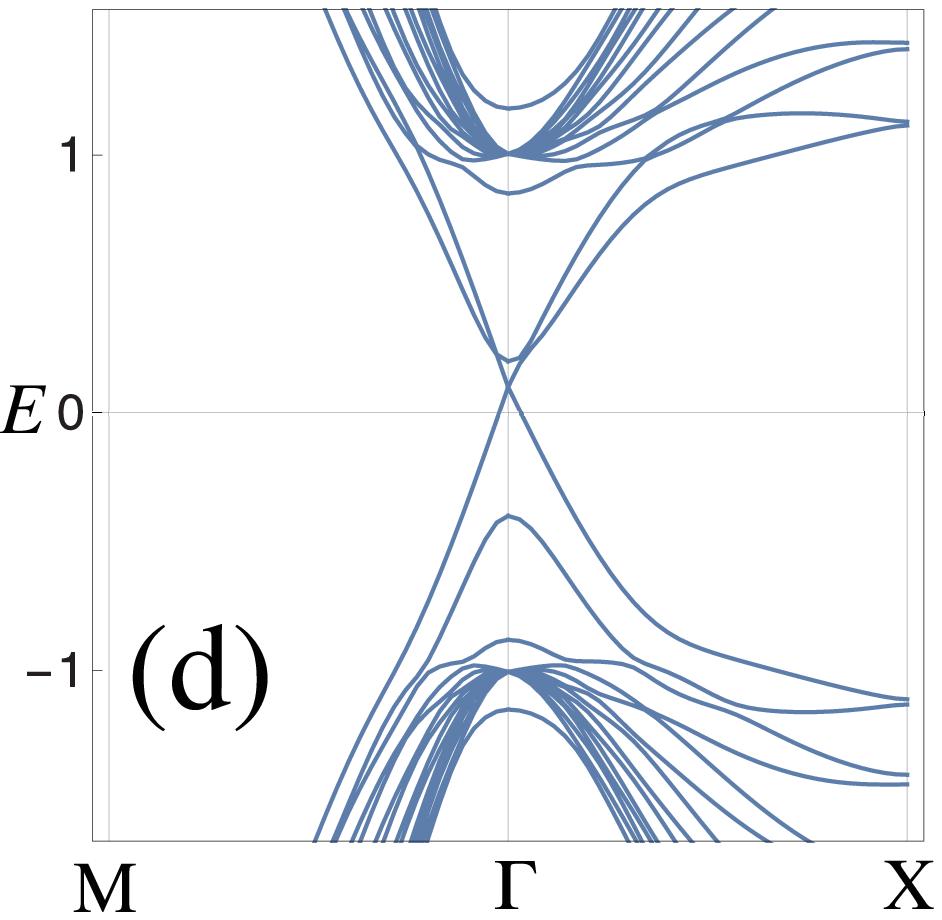}
&\includegraphics[scale=0.295]{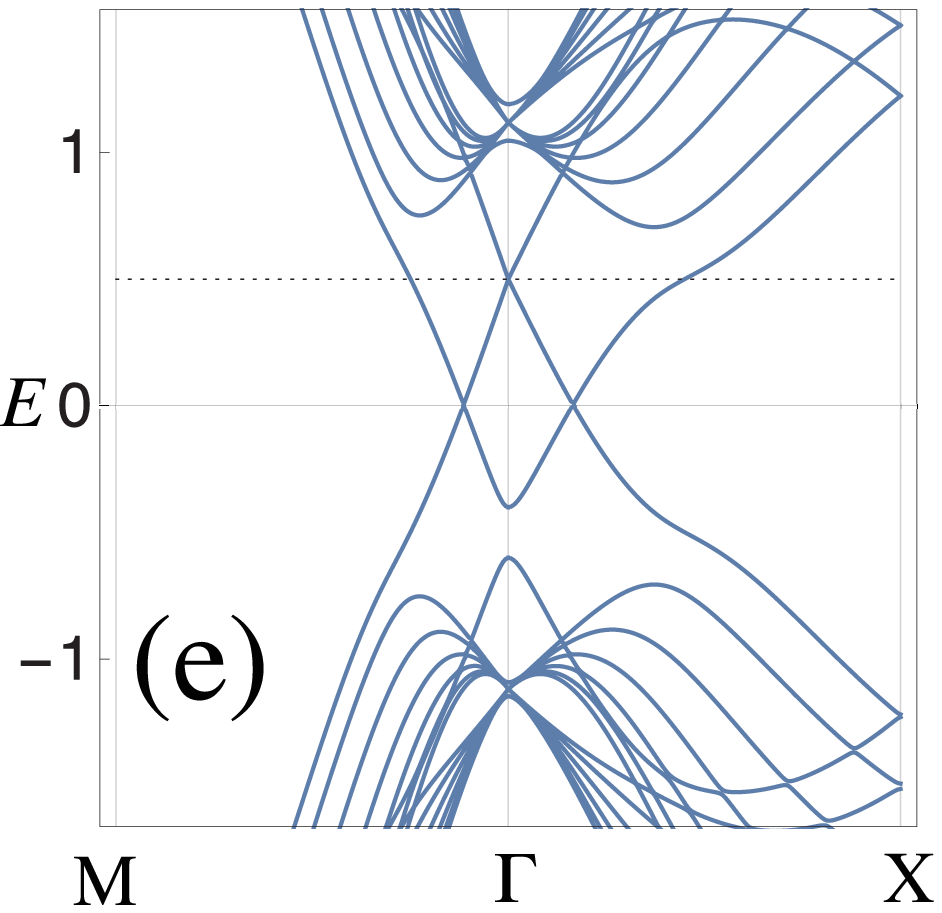}
\end{tabular}
\caption{
(a) Topological insulator with surfaces at $j_z=1$ and $j_z=n$. 
A and B show the partition for 
computing the entanglement Chern number.
(b)-(e) show the spectrum of $H$ with seven sheets ($n=7$) in various cases. 
We set $t=b=m=1$. The bulk gap opens $|E|\lesssim 1$ across the zero energy.
(b) is the spectrum of $H_0$ ($\delta H=0$). The lower three include $\delta H$ with
(c) $m_5=0.3$ and  $g=0$, (d)  $m_5=0.3$ and  $g=0.1$, and (e)
$m_5=0.1$ and  $g=0.5$. 
In (e), the Dirac point of the top surface 
is located near $E=0.5$, indicated by the dashed line.
}
\label{f:band}%-----------------------------------------------
\end{center}
\end{figure}

Let us consider the model
$H=H_0+\delta H$:  $H_0$ describes a topological insulator of Wilson-Dirac type defined by
\begin{alignat}1
H_0=&\frac{-it}{2}\sum_\mu\sum_{j}\left(c_{j}^\dagger \gamma_\mu c_{j+\hat\mu}-\mbox{h.c}\right)
+m\sum_jc_j^\dagger \gamma_4 c_j 
\nonumber\\
&+\frac{b}{2}\sum_\mu\sum_{j}\left(c_{j}^\dagger \gamma_4 c_{j+\hat\mu}+\mbox{h.c}
-2c_j^\dagger\gamma_4c_j\right),
\label{Ham0}%------- 
\end{alignat}
where we will consider a finite layered system along the $z$-direction with $j_z=1,2,\cdots,n$, and
we choose 
$\gamma_j=\tau_1\sigma_j$ ($j=1,2,3$), $\gamma_4=\tau_2$ and $\gamma_5=\tau_3$, though
any choices are possible as long as the anticommutation relation $\{\gamma_\mu,\gamma_\nu\}=2\delta_{\mu\nu}$ holds.
The two-dimensional planes 
assigned by $j_z=1$ and $n$ are called the bottom surface and the top surface, respectively,
as illustrated in Fig. \ref{f:band}(a).
Throughout the paper, the width $n$ is fixed as $n=7$ for numerical calculations. 
This model has time-reversal symmetry as well as chiral symmetry.
Therefore, we can describe the phase of the 3D bulk by two topological invariants, 
the Z$_2$ invariant respecting time reversal symmetry
and the winding number respecting chiral symmetry.\cite{Kane:2005aa,Qi:2008aa,Schnyder:2008aa}
We only consider {\it the strong topological insulator (STI) with winding number $-1$} realized in $0<m/b<2$.
The additional $\delta H$ is symmetry-breaking terms defined by
\begin{alignat}1
\delta H=&g\sum_jc_j^\dagger (i\gamma_4\gamma_3) c_j
+m_5\sum_{j}\delta_{j_z,1}c_j^\dagger \gamma_5c_j.
\label{DelHam}%-------
\end{alignat}
The former breaks chiral symmetry which we expect is originated from an intrinsic crystal structure, whereas
the latter term, breaking both symmetries, is defined only on the bottom surface 
which we assume is due to the proximity effect of an external magnetic material, etc.

Let us  first discuss the band structure of the model with particular emphasis on surface states.
In Fig. \ref{f:band}(b)-(e), we show several band diagram of $H$.
(b) is the case with $\delta H=0$, in which doubly-degenerated Dirac surface states are 
observed around the $\Gamma$-point. 
Including the symmetry-breaking $m_5$-term  at the bottom surface,
(c) shows that the degeneracy of the Dirac states is lifted:
The bottom surface state becomes massive, whereas the top surface state remains massless.
Including further the chiral symmetry breaking $g$-term into the bulk,
the spectrum becomes 
manifestly asymmetric with respect to zero energy, as seen in (d) and (e). 
Especially in case (e), chiral symmetry is so largely broken that the Dirac point of the top surface 
is almost embedded into the bulk spectrum.
In all these cases, the existence of the surface states guarantees that the 3D bulk is topological.

\begin{figure}[htb]
\begin{center}
\begin{tabular}{cc}
\multicolumn{2}{c}{\includegraphics[scale=0.75]{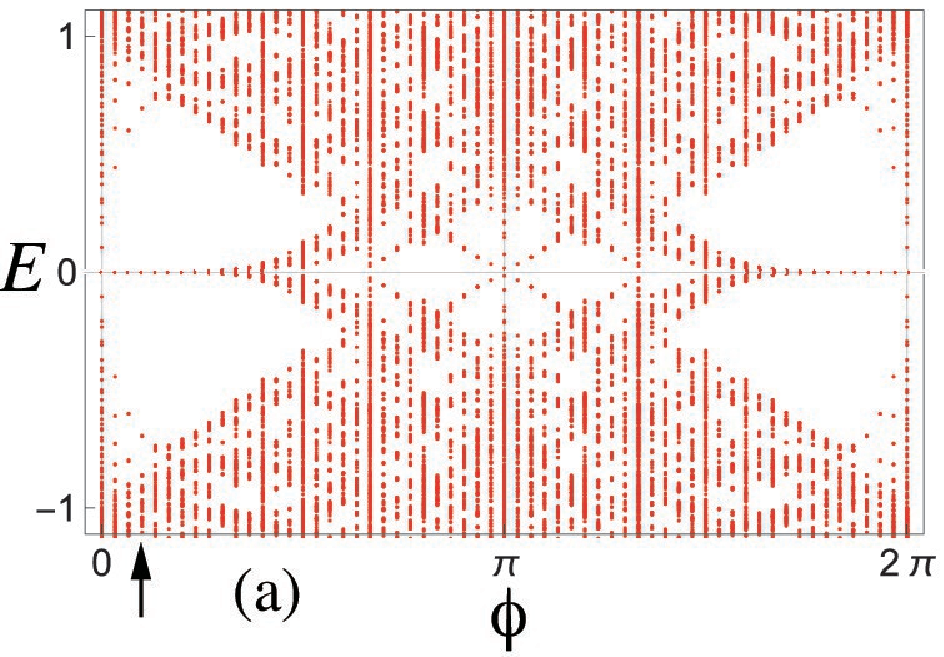}}\\
\includegraphics[scale=0.46]{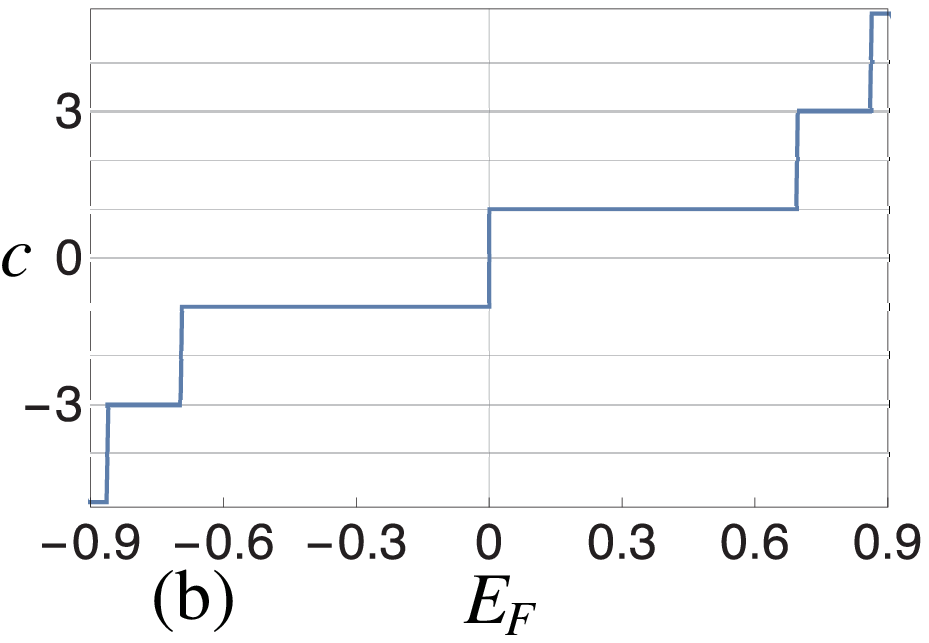}
&\includegraphics[scale=0.51]{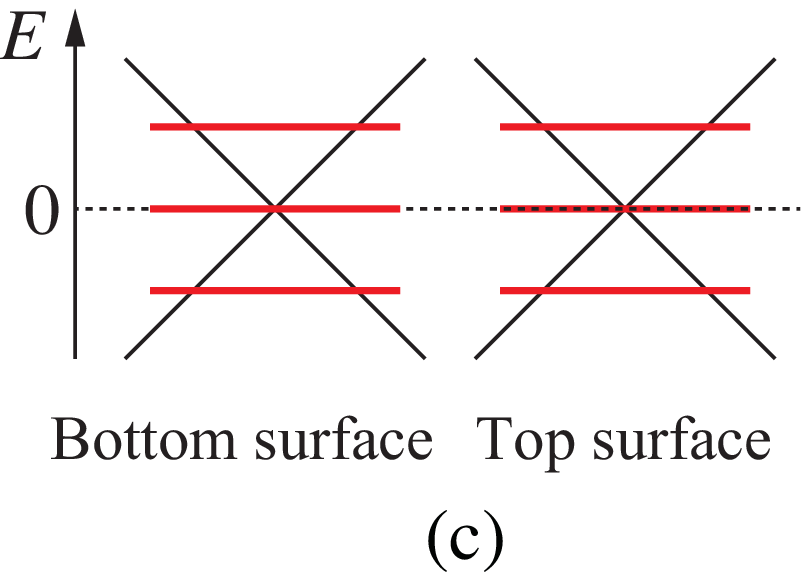}
\end{tabular}
\caption{
QHE of $H_0$ ($\delta H=0$).
The parameters
correspond to those in Fig. \ref{f:band}(b).
(a) shows a spectrum  as the function of $\phi=2\pi p/60$.
%, where $q$ is fixed as $q=60$. 
(b) shows the Chern number (the Hall conductivity in unit of $e^2/h$) 
as the function of the Fermi energy at $\phi=2\pi\cdot1/20$, indicated by an arrow in (a). 
(c) shows a schematic illustration of the Landau levels of the degenerate surface Dirac states in (b).  
}
\label{f:qhe}%-----------------------------------------------
\end{center}
\end{figure}

Now we apply a magnetic field to $H$ and investigate the QHE of the surface modes.
First, let us show the spectrum  of the model $H_0$
as a function of the magnetic flux per plaquette, $\phi=2\pi p/q$ in Fig. \ref{f:qhe}(a).
This corresponds to the famous Hofstadter butterfly.\cite{Hofstadter:1976aa}
Quantized Landau levels are observed within the bulk gap $|E|\lesssim1$ 
in a weak field regime in Fig. \ref{f:qhe} (a).
Indeed, computed Chern numbers \cite{Thouless:1982uq,kohmoto:85,FHS05} of Landau levels shows
the QHE of the Dirac fermion, as shown in (b). 
In particular, exactly degenerate zero energy Landau levels are one of the hallmarks of the Dirac fermions, 
as illustrated in (c).
The Hofstadter butterfly (a) shows that
these zero energy states are rather stable against magnetic fields. 
When a magnetic field reaches $\phi\sim2\pi/5$, the zero energy states eventually disappear. 
We expect that across this point 
the 3D bulk property may be changed from topological insulator to a trivial insulator.

\begin{figure}[htb]
\begin{center}
\begin{tabular}{cc}
\multicolumn{2}{c}{\includegraphics[scale=0.75]{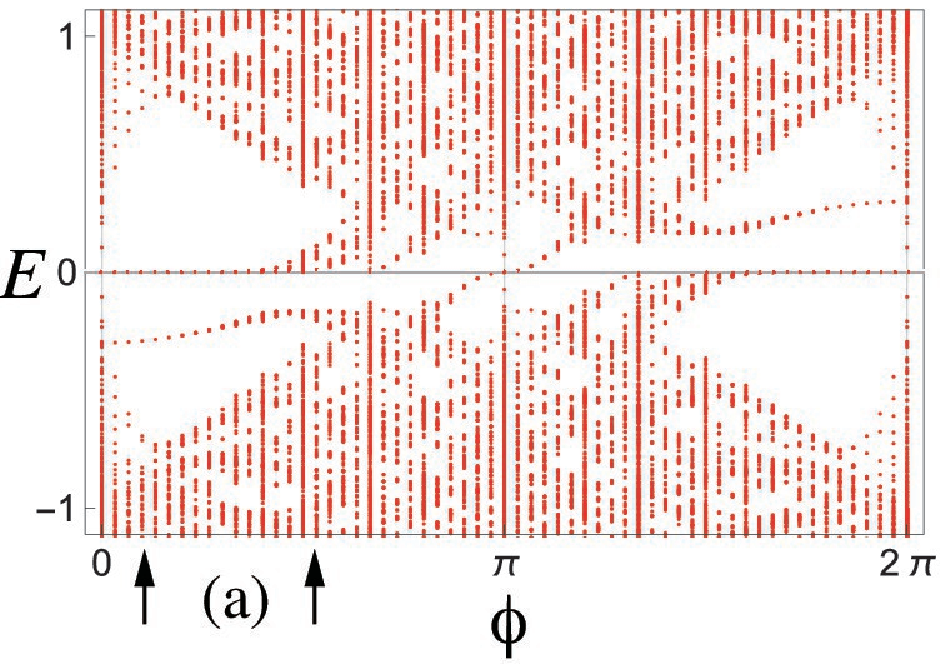}}\\
\includegraphics[scale=0.46]{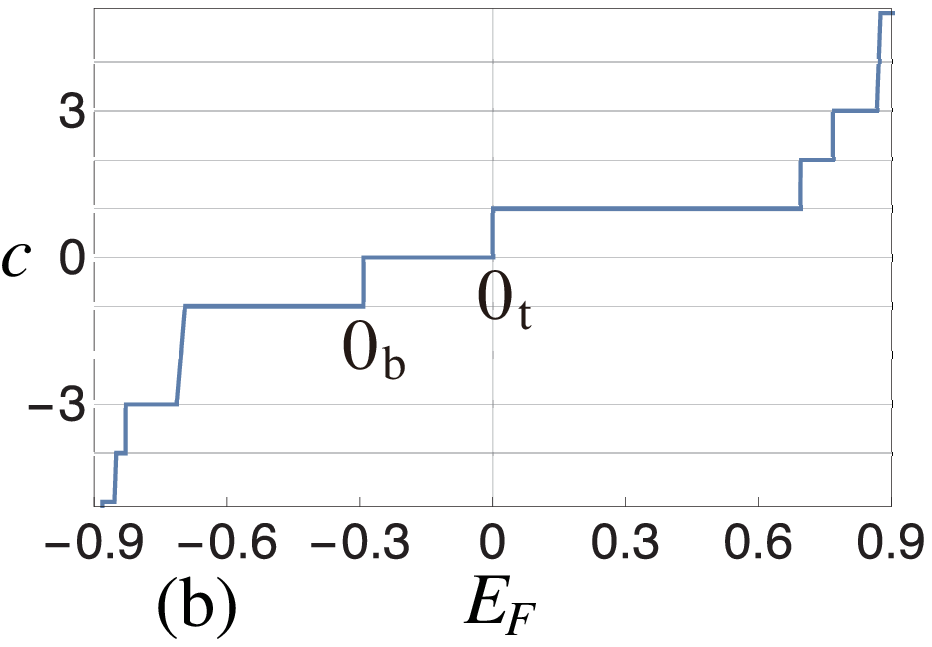}
&\includegraphics[scale=0.51]{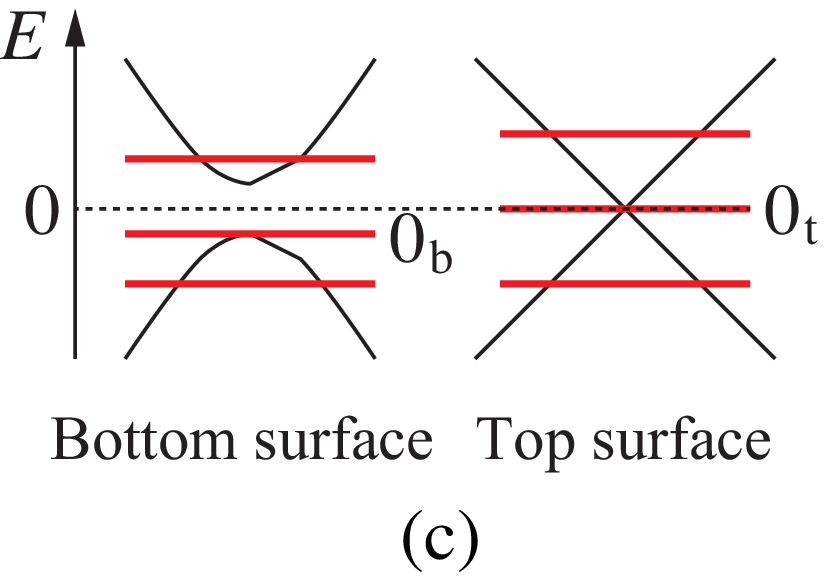}\\
\includegraphics[scale=0.46]{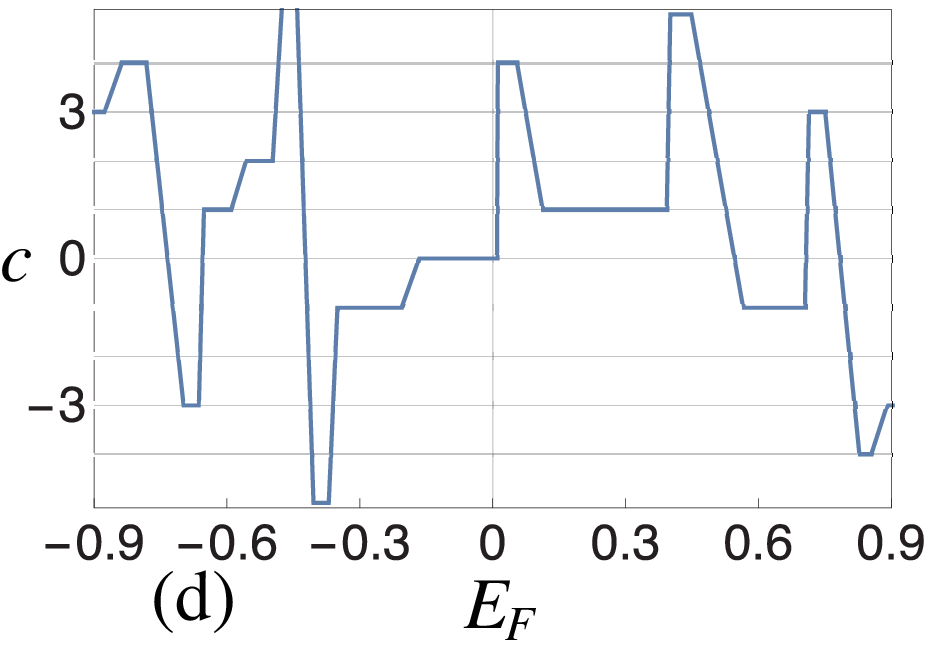}
&\includegraphics[scale=0.45]{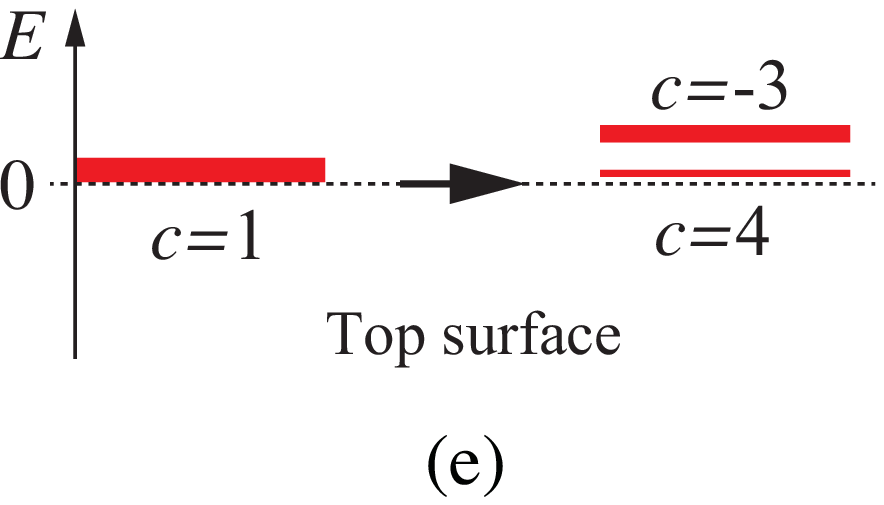}
\end{tabular}
\caption{
QHE of $H$ with $m_5$-term. 
The parameters
correspond to those in Fig. \ref{f:band}(c).
(a), (b) and (c) are similar to those in Fig. \ref{f:qhe}.
(d) shows the Chern number at $\phi=2\pi\cdot4/15$, indicated by the right arrow in (a). 
Across this point, the zero energy state disappear. 
(e) is an illustration of the Landau level splitting at this point.  
}
\label{f:qhe_m5}%-----------------------------------------------
\end{center}
\end{figure}

The model $H_0$ examined so far has chiral symmetry, which remains even under a magnetic field.
Thus, let us next study the case where chiral symmetry (as well as time reversal symmetry)
is broken at one of the surface caused by
the proximity effect of a magnetic system.
Figure \ref{f:qhe_m5} (a) shows that the degeneracy of the surface states are lifted:
The bottom surface states becomes massive and the $n=0$ Landau level,
denoted as $0_{\rm b}$ in (b) and (c), 
moves to negative energy, whereas the top surface state remains massless and yields the zero energy $n=0$
state, denoted as $0_{\rm t}$.
This zero energy state is also quite stable 
against a strong magnetic field, as shown in (a), even though a symmetry-broken proximity effect is 
taken into account.
If a magnetic field is further increased, the zero energy state becomes wider and vanishes at last.
One of interesting features here is a Landau level splitting.
As shown in (d) and (e), 
the zero energy Landau level with $c=1$ splits into two levels with $c=4$ and $c=-3$, 
and eventually move to positive energies.
At this point, the topological property of the 3D bulk may also be changed, 
which is caused by a strong magnetic field.

\begin{figure}[htb]
\begin{center}
\begin{tabular}{cc}
\multicolumn{2}{c}{\includegraphics[scale=0.75]{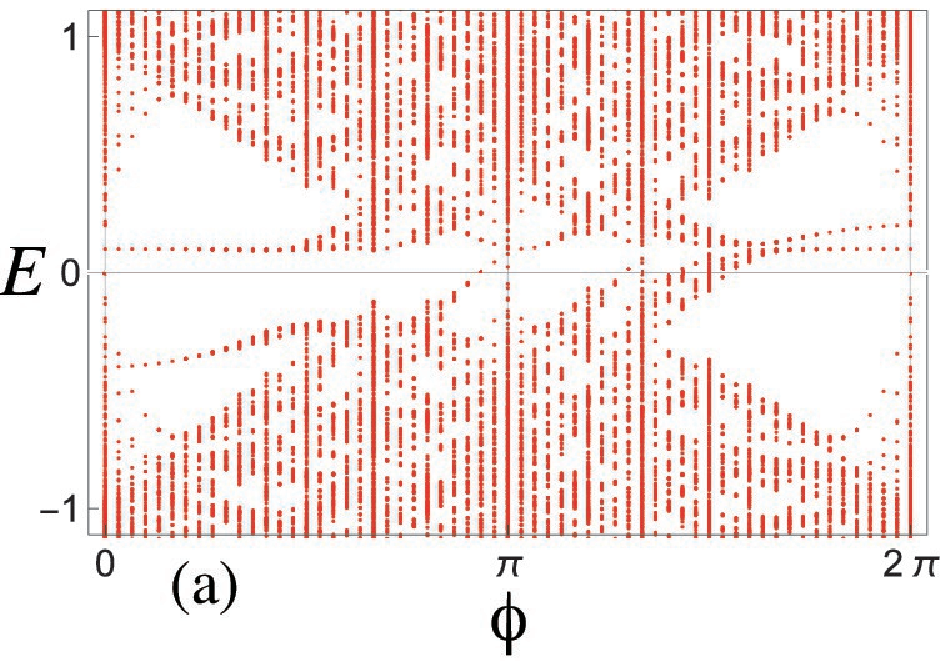}}\\
\multicolumn{2}{c}{\includegraphics[scale=0.75]{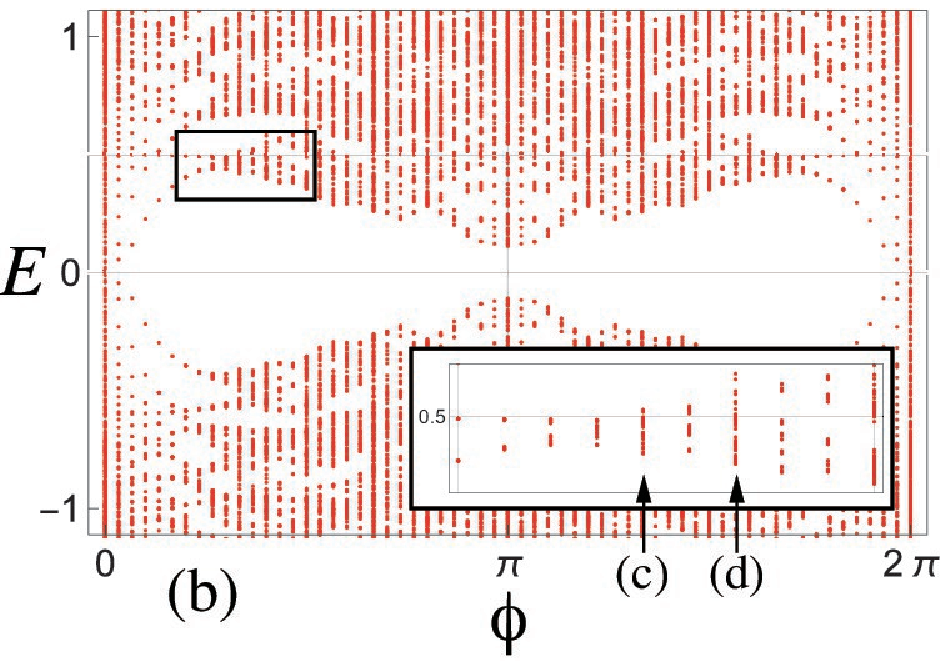}}\\
\includegraphics[scale=0.45]{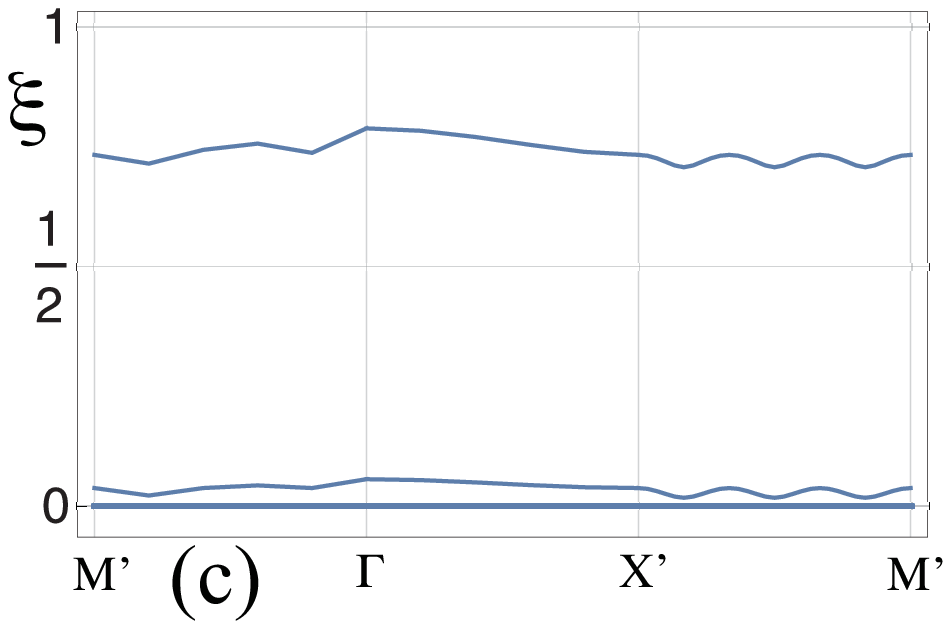}
&\includegraphics[scale=0.45]{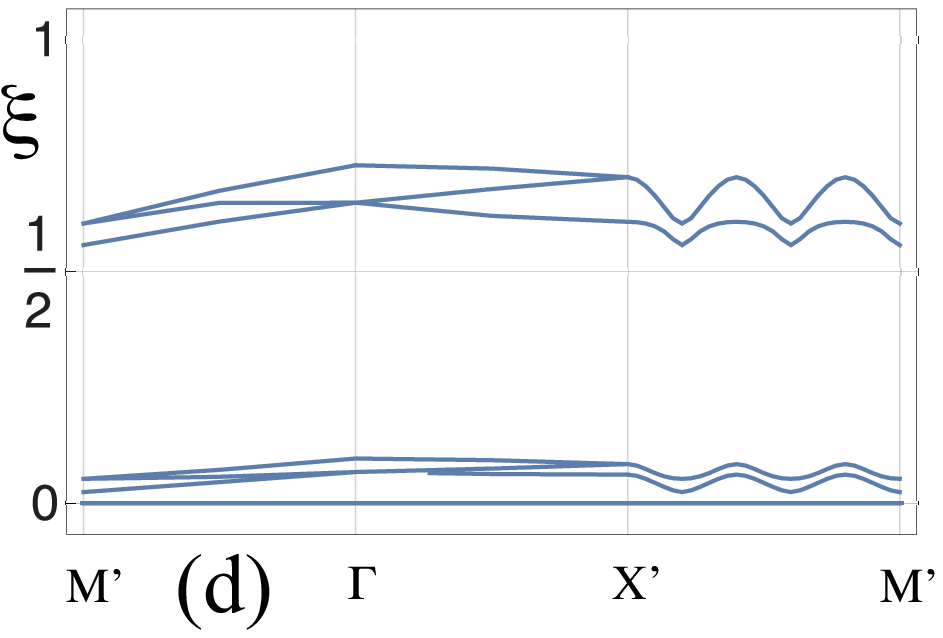}
\end{tabular}
\caption{
QHE of $H$ with $g$-term as well as $m_5$-term.
The parameters in (a) and (b) correspond to those in Fig. \ref{f:band}(d) and (e), respectively.
In (b), a horizontal line is shown at the Dirac point energy $E=0.5$ in Fig. \ref{f:band}(e), 
and the inset shows the enlarged spectrum surrounded by a rectangle.
(c) and (d) are entanglement spectra for degenerate bands including the Landau levels of the
surface Dirac modes at $\phi=2\pi\cdot1/6$ and $\phi=2\pi\cdot1/5$, indicated by arrows in (b).
$X'$ and $M'$ show $(\pi/q,0)$ and $(\pi/q,\pi)$, respectively, in the magnetic Brillouin zone 
under the Landau gauge.
}
\label{f:qhe_m5_g}%-----------------------------------------------
\end{center}
\end{figure}

Chiral symmetry seems artificial in crystals, so that we next introduce the chiral symmetry breaking $g$-term
for the bulk defined in Eq. (\ref{DelHam}).
Such a model is regarded as a generic topological insulator in STI phase, from the point of view of symmetries.
Compared with Fig. \ref{f:qhe_m5} (a), Fig. \ref{f:qhe_m5_g} (a) shows that  
the zero energy surface state is just shifted to positive energy, and is constant as a function of $\phi$. 
This is the $n=0$ Landau level of the top surface state 
located at the Dirac point which is independent of a magnetic field.
Considering the fact that the surface states of a generic topological insulator  
survive against a strong magnetic field, 
we expect that the symmetry-protected topological phase of the 3D bulk is also stable against 
a magnetic field.

\begin{figure}[htb]
\begin{center}
\includegraphics[scale=0.7]{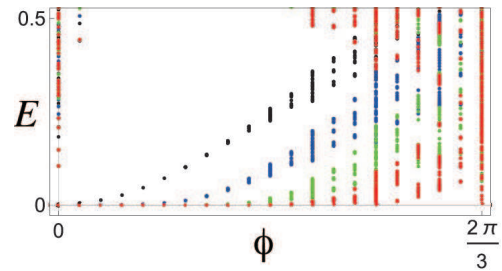}
\caption{
Hofstadter butterfly near zero energy. The parameters are those in Fig. \ref{f:band}(c) and 
Fig. \ref{f:qhe_m5} except for $m$.
Black, blue green, and red dots show the energy eigenvalues of the model with $m=0.4$, $0.6$, $0.8$, $1.0$,
respectively.  
}
\label{f:stability}%-----------------------------------------------
\end{center}
\end{figure}

Even in the extreme case in Fig. \ref{f:band}(e)
where the Dirac point is located quite near the bulk band, one can observe a level which is almost 
constant with respect to $\phi$ in a weak field regime 
at the Dirac point energy $E\sim0.5$, as shown in Fig. \ref{f:qhe_m5_g} (b).
When a magnetic field becomes stronger, as indicated by a rectangle in (b),
this Landau level with Chern number $c=1$  becomes degenerate with another level with Chern number $c=1$.
The inset of (b) shows that the two levels merge together, and as a result, they become 
a degenerate Landau level with Chern number $c=2$.
However, even within this level, the Landau level at the Dirac point keeps its character. 
To see this, let us calculate the entanglement Chern number, which has recently been introduced to 
extract a partial Chern number from degenerate bands.\cite{Fukui:2014qv,Fukui:2015fk}
Let $|\Phi\rangle$ be the occupied state of the degenerate Landau level under consideration
and let $\rho=|\Phi\rangle\langle\Phi|$ be the projection operator to this Landau level.
We consider the partition of the whole system into two subsystems, A and B, and 
integrating out B, ${\rm tr}_{\rm B}~\rho\equiv \rho_{\rm A}\propto e^{-H_{\rm A}}$, we obtain the
entanglement Hamiltonian $H_{\rm A}$.
To examine the surface state, we take A and B as the top surface and the others in Fig. \ref{f:band}(a),
respectively,
and calculate the entanglement spectrum, i.e., the eigenvalues $\epsilon_{\rm A}$ of $H_{\rm A}$.
Figures \ref{f:qhe_m5_g} (c) and (d) show the entanglement spectrum $\xi=1/(e^{\epsilon_{\rm A}}+1)$.
The bands are well-separated into two, $\xi>1/2$ and $\xi<1/2$, and 
in such a gapped case, $|\Phi\rangle$ may be adiabatically deformed 
into a single tensor product of the form\cite{Fukui:2014qv,Fukui:2015fk}
\begin{alignat}1
|\Phi\rangle\sim|\Phi\rangle_{\rm A}\otimes|\Phi\rangle_{\rm B}.
\end{alignat}
It follows that the Chern number $c$ of $|\Phi\rangle$ can be separated into 
the entanglement Chern numbers $c_{\rm A}$ and $c_{\rm B}$
associated with $|\Phi\rangle_{\rm A}$ and $|\Phi\rangle_{\rm B}$, respectively, satisfying $c=c_{\rm A}+c_{\rm B}$.
The upper bands in Fig. \ref{f:qhe_m5_g} (c) and (d) give $c_{\rm A}=1$, and hence, 
wavefunctions of the top surface (A) keep their character even if the levels merge in energy.

\begin{figure}[htb]
\begin{center}
\begin{tabular}{c}
\includegraphics[scale=0.75]{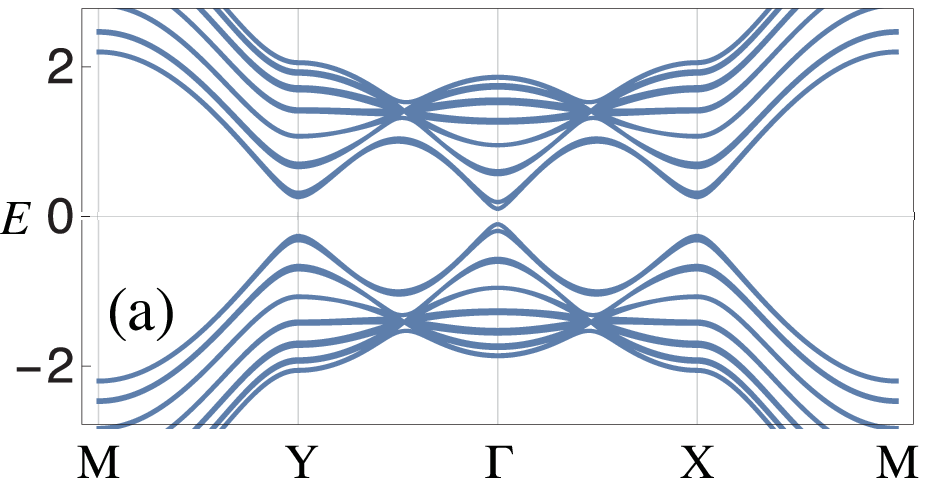}\\
\includegraphics[scale=0.75]{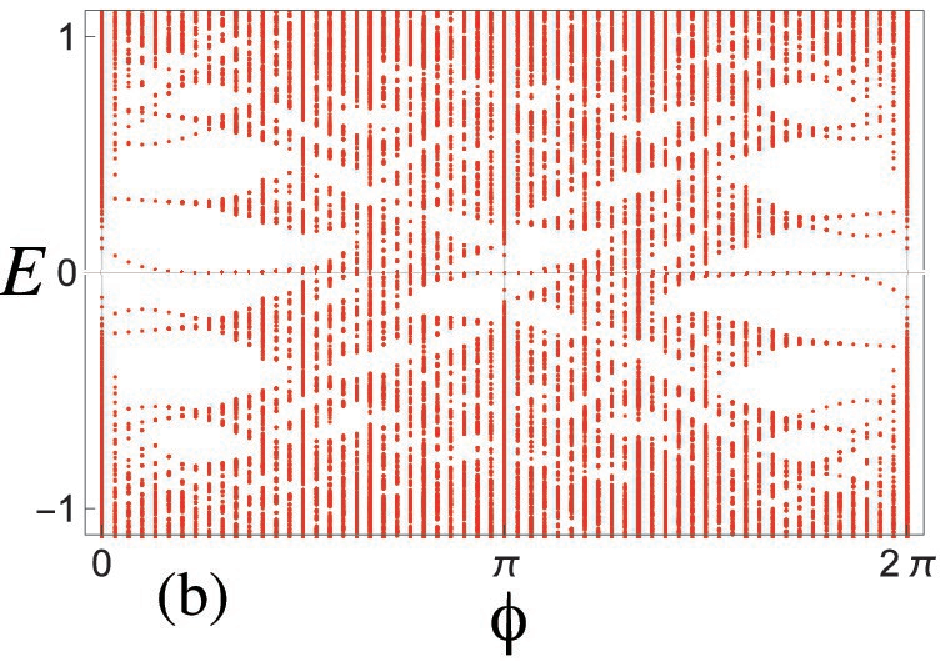}
\end{tabular}
\caption{
Model near the transition point to the WTI phase with $t=b=1$, $m=1.9$, $g=0$ and $m_5=0.3$.
(a) and (b) show the band diagram and Hofstadter butterfly, respectively.
}
\label{f:m4}%-----------------------------------------------
\end{center}
\end{figure}

So far we have consider the case with $m/b=1$. We finally discuss the effect of 
the parameter $m$ controlling the bulk gap.
At the point $m/b=0$, the bulk gap closes at the $\Gamma$ point, and for $m/b<0$ 
the groundstate becomes a trivial insulator
with winding number $0$. Let us first consider a small $m/b$ case. 
Figure \ref{f:stability} shows the spectrum near zero energy for various $m/b$.
We see that approaching the phase boundary at $m/b=0$, the zero energy states become unstable against 
a magnetic field. 
This may be natural,  since 
topological characters of the surface state become weaker.
On the other hand, at the opposite phase boundary at $m/b=2$, where bulk gap is closed at X, Y and Z points, 
and the groundstate enters {\it the weak topological insulator (WTI) phase with winding number 2} realized in 
$2<m/b<4$. 
Near the transition point, the penetration of the wave function of the surface state  becomes
deeper into the bulk, and across the transition point, the surface state switches to other surface states 
of the WTI.
Interestingly, in this case, a magnetic field can stabilize gapped surface states on
a thin film.
We show in Fig. \ref{f:m4}(a) the band structure of the model near the transition point. 
Since the penetration depth of the surface state
around the $\Gamma$ point is not negligible compared to the width of the system ($n=7$), 
the surface states on the top and bottom surfaces are hybridized, giving rise to a gap at zero energy.
Nevertheless, (b) shows that as a magnetic field becomes stronger, the gapped surface state on the top surface approaches
zero energy. This implies that a magnetic field has a tendency to stabilize surface states of topological origin.
Thus the behavior of the surface state near $m/b=0$ and $m/b=2$ is quite different,
implying that the distinct character of the surface state is revealed by a magnetic field.
Detailed analysis of surface states of the WTI under a strong magnetic field may be an interesting future problem.

It is a pleasure to thank K.-I. Imura for helpful discussions.
This work was supported by Grants-in-Aid for Scientific Research Numbers 
25400388 and 26247064
from Japan Society for the Promotion of Science.

\bibliography{s_dirac}

\end{document}